\documentclass{PoS}
\newcommand{\vareps}{\varepsilon}
\newcommand{\wt}{\widetilde}
\newcommand{\M}{\mathcal{M}}
\newcommand{\lb}{\left(}
\newcommand{\rb}{\right) }
\newcommand{\eqn}{equation}
\newcommand{\ph}{\hat{p}}
\newcommand{\al}{\alpha}
\newcommand{\ol}{\overline}
\newcommand{\as}{\alpha_{s}}
\title{Alternative subtraction scheme using Nagy Soper dipoles}

\ShortTitle{Nagy Soper substraction scheme}

\author{\speaker{Tania Robens}\\
        Department of Physics and Astronomy, University of Glasgow,
        Glasgow G12 8QQ, Scotland, UK, and\\
Institut f\"ur Theoretische Physik, RWTH Aachen University, 52056 Aachen, Germany\\
        E-mail: \email{trobens@physics.gla.ac.uk}}

\author{Cheng Han Chung\\
        Institut f\"ur Theoretische Physik, RWTH Aachen University, 52056 Aachen, Germany\\
        E-mail: \email{chenghan@physik.rwth-aachen.de}}

\abstract{We present an alternative subtraction scheme for the treatment of infrared divergences in NLO QCD
calculations. In this scheme, the number of transformations is greatly reduced with respect to the
 standard subtraction scheme by Catani and Seymour. We discuss the general setup of the scheme as
well as first applications to NLO processes at hadron and lepton colliders.\\
TTK-10-07
}

\FullConference{RADCOR 2009 - 9th International Symposium on Radiative Corrections (Applications of Quantum Field Theory to Phenomenology) ,\\
		 October 25 - 30 2009\\
		 Ascona, Switzerland}

\begin{document}

\section{Introduction: General structure of subtraction schemes}
In higher order calculations, the cancellation of infrared singularities
is generally treated by the introduction of an infinitesimal regulator, eg in the form of a finite
mass for the massless gauge boson, or by lowering the dimension $D$ of the
respective phase space integrals to $D\,=\,4-2\,\vareps$. This way,
the analytic
cancellation of the respective divergent terms for fully inclusive
variables after phase space integration is straightforward; however, numerical implementations of
terms containing small regulators prove to be challenging. In subtraction schemes, this
problem is circumvented by the introduction of local counterterms,
which mimic the behaviour of the squared real emission matrix elements in the
singular regions; adding back the respective one particle integrated counterparts to the
virtual contributions results in finite integrands for both real
emission and virtual contribution phase space. Symbolically, this is
given by
\begin{eqnarray}
\sigma^{\rm{NLO}}&=&\sigma^{\rm{Born}}\,+\,\sigma^{\rm{virt}}\,+\,\sigma^{\rm{real}}
\,=\,\sigma^{\rm{Born}}\,+\,\sigma^{\rm{virt}}\,+\,\sigma^{\wt{A}}\,+\,\sigma^{\rm{real}}\,-\,\sigma^{A},
\end{eqnarray}
where
\begin{eqnarray}\label{eq:gen_expl}
\sigma^{\rm{Born}}\,+\,\sigma^{\rm{virt}}\,+\,\sigma^{\wt{A}}&=&\int\,d\Gamma_{m}\left[
|\M_{\rm{Born}}|^{2}\,+\,2\,\rm{Re}\lb\M_{Born}\M_{\rm{virt}}^{*}\rb\,+\,\sum_{i}\mathcal{V}_{i}|\M_{\rm{Born}}|^{2}
\right],\nonumber\\
\sigma^{\rm{real}}\,-\,\sigma^{A}&=&\int\,d\Gamma_{m+1}\,\left[|\M_{\rm{real}}|^{2}\,-\,\sum_{i}D_{i}|\M_{\rm{Born}}|^{2}\right]
\end{eqnarray}
are the respective $m,\,m+1$ phase space contributions to the total NLO
cross section\footnote{For hadronic initial states, an additional collinear
counterterm $\sigma^{C}$ needs to be added in $m$ particle phase
space, which accounts for contributions already contained in the NLO
PDFs.}. Convolution with jet functions then allows to define
differential quanities and guarantees infrared safety
of the respective Born contribution.
In eq (\ref{eq:gen_expl}), the sum goes over all local counterterms needed to
match the complete singularity structure of the real emission contribution. 
For each singular limit, the real emission
matrix element factorizes according to
\begin{\eqn}\label{eq:fac}
|\M_{\rm{real}}|^{2}(p)\,\longrightarrow\,D\,|\M_{\rm{Born}}|^{2}(\tilde{p}),
\end{\eqn}
where $D$ denotes the dipole containing the respective singularity structure.
As $\M_{\rm{real}}$ and $\M_{\rm{Born}}$ live in different
phase spaces, a mapping of the respective momenta from $m+1$ to $m$
particle phase space needs to be introduced, which is defined by a
mapping function $F_{\rm{map}}$ according to
$\tilde{p}\,=\,F_{\rm{map}}\,(p).$
 While the complete singularity
structure of the limit considered is
 contained in $D$, both $D$
and $\M_{\rm{Born}}$ can depend on the leftover nonsingular parameters
of phase space. $D_{i}$ and $\mathcal{V}_{i}$ are related
by 
$\mathcal{V}_{i}\,=\,\mu^{2\,\vareps}\int\,d\xi\,D_{i}$,
where the integration measure $d\xi$ is an effective one particle integral.\\ 
Summarizing, any subtraction scheme needs to fulfill 
the following requirements:
\begin{itemize}
\item{}definition of subtraction terms $D_{i}$ which,
  following eq (\ref{eq:fac}), one by one mimic the
  behaviour of the real emission matrix element in each singular
  region such that their sum contains the complete singularity
  structure of the process,
\item{}definition of a mapping $F_{\rm{map}}$ which guarantees total energy
  momentum conservation as well as onshellness of all external
  particles both before and after the mapping,
\item{}integration measure $d\xi$, with a ``smart'' choice of variables
  providing optimal singularity structure parametrization.
\end{itemize}
In the following, we will discuss this for our specific scheme,
comparing with \cite{Catani:1996vz} where appropriate. 
\section{Scheme setup}
In the scheme discussed in this report, the NLO subtraction terms are
derived from the splitting functions introduced in \cite{Nagy:2007ty}, and the
$m+1$ to $m$
phase space mappings needed correspond to the inverse of the
respective shower $m$ to $m+1$ mappings. In the following, we will
denote the $m+1$ phase space four vectors by $\ph_{1},\ph_{2},...$ and
$m$ phase space four vectors by $p_{1},\,p_{2},\,...$. Indices $a,\,b$
will denote initial state particles; in $m+1$ phase space, $\ph_{i}$
is the emitter, $\ph_{j}$ the emitted particle, and $\ph_{k}$ the
spectator\footnote{In contrast to \cite{Catani:1996vz}, in our case a
  spectator only needs to be specified if $\ph_{j}$ denotes a
  gluon.}. By default, for initial state emissions we set $\ph_{i}\,=\,\ph_{a}$ in all general
expressions below. Equally, we restrict our expressions to subtractions on the parton level and to the massless case; details on convolution with PDFs are given in \cite{Nagy:2007ty}, and the extension to massive particles is in the line of future work.
\subsection{Momentum mapping and integration measure: Initial state}
For the initial state, the mapping from the $m+1$ to the $m$ particle phase space is given by
\begin{eqnarray}
&&p_{a}\,=\,\lb 1-\frac{\ph_{j}\cdot\hat{Q}}{\ph_{a}\cdot\ph_{b}} \rb\,\ph_{a},\;p_{l}\,=\,\Lambda ( K,\widehat{K} )\,\ph_{l},\,p_{b}\,=\,\ph_{b},
\end{eqnarray}
where the index $l$ goes over all final state particles in the $m$ particle phase space and with
\begin{eqnarray}\label{eq:LTini}
&&\Lambda (K,\widehat{K})\,=\,g\,-\,\frac{2\,(K+\widehat{K})\,(K+\widehat{K})}{(K+\widehat{K})^{2}}\,+\,\frac{2\,K\,\widehat{K}}{\widehat{K}^{2}},
\end{eqnarray}
where $K\,=\,p_{a}+p_{b},\,\hat{Q}\,=\,\ph_{a}\,+\,\ph_{b},\; \widehat{K}\,=\,\hat{Q}\,-\,\ph_{j}$.
The phase space factorizes according to
\begin{\eqn}
\left[d\{\ph,\hat{f}\}_{m+1}\right]g(\{\ph,\hat{f}\}_{m+1})\,=\,\left[d\{p,f\}_{m}\right]\,d\xi_{p}g(\{\ph,\hat{f}\}_{m+1}),
\end{\eqn}
where $f$ denotes the flavour and with the D-dimensional integration measure
\begin{\eqn}
d\xi_{p}\,=\,\frac{d^{D}\ph_{j}}{(2\,\pi)^{D-1}}\,\delta_{+}\lb\ph_{j}^{2}\rb.
\end{\eqn}
\subsection{Momentum mapping and integration measure: Final state}
For final state splittings, the initial state momenta remain unchanged:
$p_{a}\,=\,\ph_{a},\;p_{b}\,=\,\ph_{b}$.
The mapping uses all non-emitting particles as one spectator for
momentum redistribution. We introduce the additional variables
\begin{eqnarray}\label{eq:vars_fin}
&&P\,=\,\ph_{i}\,+\,\ph_{j},\;Q\,=\,\ph_{a}+\ph_{b},\nonumber\\
&&y\,=\,\frac{P^{2}}{2\,P\cdot Q-P^{2}},\;a\,=\,\frac{Q^{2}}{2\,P\cdot Q-P^{2}},\;
\lambda\,=\,\sqrt{(1+y)^{2}-4\,a\,y}.
\end{eqnarray}
The emitting particle is mapped according to
\begin{\eqn}\label{eq:pimap_fin}
p_{i}\,=\,\frac{P}{\lambda}\,-\,\frac{1-\lambda+y}{2\,\lambda\,a}\,Q.
\end{\eqn}
All non-emitting final state particles are mapped using the Lorentz transformation as in eq. (\ref{eq:LTini}), where now
$K\,=\,Q-p_{i},\;\widehat{K}\,=\,Q-P.$
Especially, this means that the total number of mappings needed for a
$N$-jet final state scales as $\frac{N^{2}}{2}$, which reduces scaling\footnote{The subtraction scheme in \cite{Frixione:1995ms} has a scaling similar to our scheme.}
 with respect to \cite{Catani:1996vz} by a factor $N$.\\
The phase space factorization takes a similar form as in the initial state splitting, ie we have again
\begin{\eqn}
\left[d\{\ph,\hat{f}\}_{m+1}\right]g(\{\ph,\hat{f}\}_{m+1})\,=\,\left[d\{p,f\}_{m}\right]\,d\xi_{p}g(\{\ph,\hat{f}\}_{m+1})\nonumber
\end{\eqn}
where now 
\begin{eqnarray}\label{eq:meas_fin}
d\xi_{p}&=&dy\,\theta(y_{\rm{max}}-y) \,\lambda^{D-3}\,\frac{p_{i}\cdot\,Q}{\pi}\,\frac{d^{D}\ph_{i}}{(2\,\pi)^{D-1}}\,\delta_{+}\,(\ph_{i}^{2})\,\frac{d^{D}\ph_{j}}{(2\,\pi)^{D-1}}\,\delta_{+}\,(\ph_{j}^{2})\nonumber\\
&&\times\,(2\,\pi)^{D}\,\delta^{(D)}\,\lb P-\lambda\,p_{i}\,-\,\frac{1-\lambda+y}{2\,a}\,Q\rb.
\end{eqnarray}
$y_{\rm{max}}\,=\,\lb\sqrt{a}-\sqrt{a-1} \rb^{2}$ can directly be derived from total energy momentum conservation.
\subsection{Treatment of interference terms: dipole partitioning functions}
Double poles in splitting functions only arise if the emitted particle
is a gluon; in this case, interference terms between different
emitters have to be taken into account. In our scheme, we split the
collinear and soft parts of the respective spin averaged splitting
functions $\ol{W}$ according to \cite{Nagy:2008ns}
\begin{\eqn}
\ol{W}_{ii}\,-\,\ol{W}_{ik}\,=\,\lb \ol{W}_{ii}\,-\,\ol{W}^{\rm{eik}}_{ii}  \rb \,+\,\lb \ol{W}^{\rm{eik}}_{ii}\,-\,\ol{W}_{ik}  \rb,
\end{\eqn}
where $\ol{W}^{\rm{eik}}_{ii}$ is the spin-averaged eikonal
factor. The second part of the above equation can be then expressed in
terms of dipole partitioning functions $A'_{ik}$ \cite{Nagy:2008eq}
\begin{\eqn}\label{eq:dip_part}
 \ol{W}^{\rm{eik}}_{ii}\,-\,\ol{W}_{ik} \,=\,4\,\pi\,\al_{s}\,A'_{ik}\,\frac{-\hat{P}_{ik}^{2}}{(\ph_{j} \cdot\ph_{i}\,\ph_{j}\cdot\,\ph_{k})^{2}},
\end{\eqn}
where $\hat{P}_{ik}\,=\,\ph_{j}\cdot\,\ph_{i}\,\ph_{k}\,-\,\ph_{j}\cdot\ph_{k}\,\ph_{i}$. Several choices for $A_{ik}'$ have been proposed \cite{Nagy:2008eq}; all results given here have been obtained using eq (7.12) therein.
\section{Example of integrated splitting function: $g\,\rightarrow\,q\,\bar{q}$ final state splitting}
For our scheme, all collinear as well as singular parts of the soft splitting
functions have been tested; a complete list will be given in \cite{future}. In this section, we give the final state
$g\,\rightarrow\,q\,\bar{q}$ dipole and the corresponding integrated
term as an example, additionally commenting on the limit for
$m\,\rightarrow\,2$.\\
\\
For a $g\,\rightarrow\,q\,\bar{q}$ splitting in the massless case, the spin averaged subtraction term is given by $D_{g q\bar{q}}|\M_{\rm{Born}}(p)|^{2}$, with
\begin{\eqn}
D_{g q\bar{q}}\,=\,T_{R}\,\frac{4\,\pi\,\al_{s}}{y\,p_{i}\cdot\,Q}\,\left[1-\frac{2\,z\,(1-z)}{1-\vareps} \right],
\end{\eqn}
where we introduced the additional variables 
$z\,=\,\frac{\ph_{j}\tilde{n}}{P\tilde{n}},\,\tilde{n}=\frac{1+y+\lambda}{2\lambda}Q-\frac{a}{\lambda}\,P$,
and all other variables as in eq. (\ref{eq:vars_fin}). This
subtraction term was derived by squaring the respective final state
shower splitting function in \cite{Nagy:2007ty}. Momentum mapping is
done according to eq. (\ref{eq:pimap_fin}) and the Lorentz transform
with the respective expressions for $K,\,\hat{K}$. For the integrated splitting function, we rewrite the measure (\ref{eq:meas_fin}) in terms of the variables introduced above and obtain
\begin{\eqn}
d\xi_{p}\,=\,{\frac{{(2\,p_{i}\,Q)^{1-\vareps}}}{16\,\pi^{2}}\,\frac{d\Omega_{d-2}}{(2\,\pi)^{1-2\,\vareps}}\,dz\,dy\,{\lambda^{1-2\,\vareps}}y^{-\vareps}\,\left[z\,(1-z)\right]^{-\vareps}}\,\theta(y_{\rm{max}}-y)\,\theta\left[ z\,(1-z)\right]
\end{\eqn} 
which results in
\begin{\eqn}
\mathcal{V}\,=\,\mu^{2\,\vareps}\,\int\,d\xi_{p}\,D\,=\,{ T_{R}\,\frac{\al_{s}}{2\,\pi}\,\frac{1}{\Gamma(1-\vareps)}\lb\frac{2\,\pi\,\mu^{2}}{ p_{i}Q}\rb^{\vareps}\,}
{\left[-\frac{2}{3\,\vareps}\,-\,\frac{16}{9}\,+\,{\frac{2}{3}\,\left[(a-1)\,\ln(a-1)-a\,\ln\,a\right]}\right]}.
\end{\eqn}
As expected, for $m\,=\,2$ the above expressions as well as the mapping completely
reduce to the result in \cite{Catani:1996vz}.\\
 We want to comment that
in our scheme, the most complicated expressions stem from the
integration of the interference terms as in eq. (\ref{eq:dip_part}). As all final state particles are mapped using the Lorentz transform $\Lambda$, the finite parts of the respective subtraction functions need to be evaluated numerically; details will be given in \cite{future}.
\section{First results}
As an example, we give the analytic result of our splitting functions
when applied to dijet production at lepton colliders, as well as a
numerical comparison for Drell-Yan process at NLO using
\cite{Catani:1996vz} as well as the scheme proposed in this writeup. We additionally tested the scheme for Higgs production at hadron colliders and decay; the respective calculations will be presented elsewhere.
\subsection{Dijet production at lepton colliders}
For dijet production at lepton colliders, the final state squared
splitting function $D\,=\,D_{qqg}$ is needed. We denote the
four-momenta of the outgoing partons in this process with
$\ph_{1}(q),\,\ph_{2}(\bar{q}),$ and $\ph_{3}(g)$. The unintegrated dipole
subtraction term for emission from $\ph_{1}$ is then given by
\begin{eqnarray}
D &=&
\frac{8\,\pi\,\al_{s}}{\hat Q^2}\,C_{F}\, \left\{ \left(\frac{1}{x_2}\right)
\left[2\left(\frac{x_1}{2-x_1-x_2}- \frac{1-x_2}{(2-x_1-x_2)^2}\right)
+\frac{1-x_1}{1-x_2}\right]  \right. \nonumber\\
&+& \left. 2 \left( \frac{x_1+x_2-1}{1-x_2} \right)\frac{x_1}{(1-x_1)x_1+(1-x_2)x_2}\right\}
\end{eqnarray}
with $x_{n}\,=\,\frac{2\,\ph_{n}\,\hat{Q}}{\hat{Q}^{2}}$. The
respective integrated averaged splitting function is
\begin{\eqn}
\mathcal{V}\,=\,\frac{\alpha_{s}}{2\pi}C_F
\frac{1}{\Gamma(1-\vareps)}\left(\frac{4\pi\mu^2}
{ Q^2}\right)^\vareps\,\left[\frac{1}{\vareps^{2}}+\frac{3}{2\,\vareps}\,-\,1\,+\,\frac{\pi^{2}}{6}\right].
\end{\eqn} 
Combining the above splitting functions for both emitters with the
Born, real emission, and virtual matrix elements and integrating over phase space, we obtain the standard result
\begin{eqnarray}
\sigma^{NLO}&=&\sigma^{NLO\,\{2\}}+\sigma^{NLO\,\{3\}}\,=\,\frac{\as}{2\pi}C_F 
\left[ \left(-10+\frac{4}{3}\pi^2\right)\,+\, \left(\frac{23}{2}-\frac{4}{3}\pi^2\right)\right]
\sigma^{LO}\,=\,  \frac{3}{4} \frac{\as}{\pi}C_F 
\sigma^{LO}.\nonumber\\
&&
\end{eqnarray}
\subsection{Drell-Yan production}
\begin{figure}
\centering
\includegraphics[width=2.8in]{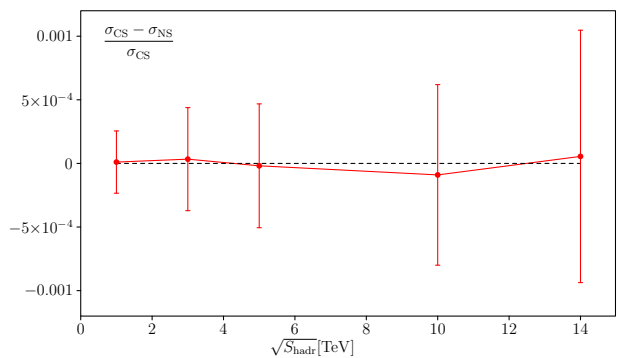}
\vspace{5mm}
\caption{\label{fig:diffdy} Relative difference between NLO corrections to single W production using Catani Seymour and Nagy Soper dipoles respectively, as a function of the hardonic cm energy. The results agree on sub-permil level, shown are the numerical integration errors. }
\end{figure}

We calculated single $W$ production for a $pp$ initial state at NLO, using both the scheme in \cite{Catani:1996vz} as well as our scheme, including PDFs and varying the hadronic cm energy of the process.
We here only show the numerical result for this process. Figure \ref{fig:diffdy} plots the relative difference between
the two implemented schemes. We see that the numerical differences are
on the permill level and consistent with zero.
\section{Summary and Outlook}
In this report, we introduce a subtraction scheme which reduces the number of mappings in the real emission part of an NLO calculations by a factor $N_{\rm{jets}}$ with respect to the scheme suggested in \cite{Catani:1996vz}. We explained the setup as well as phase space mapping, and presented first results for an integrated splitting function as well as applications for simple processes. A complete listing of all integrated splitting functions as well as further examples will be given in \cite{future}.

\end{document}